\documentclass[pre,aps,floatfix,twocolumn,superscriptaddress,showpacs]{revtex4}
\usepackage{color}
\usepackage{graphicx}
\usepackage{amsmath}
\usepackage{amsthm}
\usepackage{amssymb}
\usepackage{lipsum}
\usepackage{latexsym}

\newcommand{\be}{\begin{equation}}
\newcommand{\ee}{\end{equation}}
\definecolor{lightblue}{rgb}{.0, .0, .8}

\graphicspath{{./figures/}}

\begin{document}
\date{\today}
\title{Hydrodynamic correlations in shear flow: A Multiparticle--Collision--Dynamics simulation study}
\author{Anoop Varghese, Chien-Cheng Huang, Roland G. Winkler, and Gerhard Gompper}
%\email{a.varghese@fz-juelich.de}
\affiliation{Institute of Complex Systems and Institute for Advanced
Simulation, Forschungszentrum J\"{u}lich, J\"{u}lich 52425, Germany}

\begin{abstract}
The nonequilibrium hydrodynamic correlations of a Multiparticle-Collision-Dynamics (MPC) fluid in shear flow are studied by analytical calculations and simulations. The Navier-Stokes equations for a MPC fluid are linearized about the shear flow and the hydrodynamic modes are evaluated as an expansion
in the momentum vector. The shear-rate dependence and anisotropy of the transverse and longitudinal velocity correlations are analyzed. We demonstrate that hydrodynamic correlations in shear flow are anisotropic, specifically,  the two transverse modes are no longer identical. In addition, our simulations reveal the directional dependence of the frequency and attenuation of the longitudinal velocity correlation function.
Furthermore, the velocity autocorrelation functions of a tagged fluid particle in shear flow are determined.
The simulations results for various hydrodynamic correlations agree very well with the theoretical predictions.

\end{abstract}
\date{\today}
\maketitle
\section{Introduction}
The thermodynamics of systems far from equilibrium has drawn growing interest
in the last couple of decades~\cite{seifert2012}. Several non-equilibrium relations, collectively called fluctuation relations,
have been derived for transient and steady non-equilibrium states. These relations have been verified using  exactly solvable models and numerical
simulations (see Ref.~\cite{seifert2012} and references therein).
An interesting class of non-equilibrium systems is fluids under external fields such as shear flow and/or a temperature gradient~\cite{zarate_book}.
Considerable progress has been achieved in understanding these systems using hydrodynamics calculations~\cite{lutsko1985,machta1980,wada2003,zarate2004,zarate2008,otsuki2009-epj},
numerical simulations~\cite{otsuki2009-pre,otsuki2009-jsm,belushkin2011}, and experiments~\cite{segre1992}.
For instance, the fluctuation relation for entropy production has been verified in numerical simulations of simple fluids under shear flow \cite{evans1993,belushkin2011}.
Apart from satisfying fluctuation relations, non-equilibrium fluids show several interesting features which are absent in equilibrium.
In particular, non-equilibrium  hydrodynamic correlations in steady states are long-ranged even for fluids far from critical points
~{\cite{machta1980,lutsko1985,lutsko2002,zarate2008,otsuki2009-pre}}.
In addition, these correlations are anisotropic, in contrast to equilibrium correlations in simple fluids.
A consequence of the long-range nature of the correlations is the non-intensivity of pressure fluctuations~\cite{wada2003}.

Computer simulations are extremely valuable to study nonequilibrium phenomena. In particular, recently developed mesoscale hydrodynamic simulations approaches, such as lattice Boltzmann~\cite{mcna:88,shan:93,he:97}, dissipative particle dynamics (DPD)~\cite{hoog:92,espa:95,espa:95.2}, or multiparticle collision dynamics (MPC) \cite{male:99,kapr:08,gomp:09}, permit to cover large length and long time scales, and a wide range of external parameters such as shear rates and temperature gradients. All the approaches are essentially alternative ways of solving the Navier-Stokes equations for the fluid dynamics. Common to them is a simplified, coarse-grained description of the fluid degrees of freedom while maintaining the essential microscopic physics on the length scales of interest \cite{gomp:09}. By now, the MPC method has successfully been applied in a broad range of equilibrium and nonequilibrium simulations of soft matter systems (see, e.g., Ref.~\cite{huan:15} and references therein). In particular,  the hydrodynamic correlations of the MPC fluid have been determined and it has been shown that they agree with the solutions of the fluctuating Landau-Lifshitz Navier-Stokes equations \cite{huang2012}. Moreover, the hydrodynamic correlations of embedded colloids \cite{lee:04,padd:05,goet:10,whit:10,belu:11,pobl:14} and polymers \cite{huan:13} have been calculated. Even more, MPC simulations have  been successfully applied to verify the fluctuation relation for entropy production in shear flows \cite{belushkin2011}. So far however, an analysis of nonequilibrium correlation functions of a MPC fluid and a comparison with theoretical approaches is missing.

In this paper, we fill this gap and determine analytically and by MPC simulations the time-correlation functions of hydrodynamic variables of a simple isothermal fluid under shear flow.
We first derive analytical expressions for the respective  correlations by linearizing the Navier-Stokes equations.
To this end, we follow the methods employed in Refs.~\cite{lutsko1985,otsuki2009-epj}, where adiabatic or granular fluid are considered. Here, the isothermal approach is simpler, because energy is no longer a conserved quantity.
We restrict ourselves to moderate shear rates for which the coupling between hydrodynamic modes can be ignored~\cite{lutsko1985,otsuki2009-jsm}.
Exploiting the MPC method, we then perform shear flow simulations and calculate the respective hydrodynamic correlation functions.
The primary effect of shear is the anisotropy of the hydrodynamic correlation functions, as already predicted in Refs.~\cite{lutsko1985,otsuki2009-jsm}.
The frequency and attenuation of the longitudinal modes become directional  and shear rate dependent.
In addition, the degeneracy of the two transverse modes, present at equilibrium, is removed.
The anisotropy of the longitudinal and transverse velocity autocorrelations is also manifested in the anisotropy of the velocity autocorrelations of tagged MPCs particles.
Moreover, the correlation functions show a faster decay than the equilibrium correlations at long times.
By comparison, we find excellent agreement between the theoretical predictions and the MPC simulation results.

The article is organized as follows. The theoretical expressions for the velocity correlation functions are derived in Sec.~\ref{theory}.
Section~\ref{sec:simulations} presents simulation results and a comparison with the theoretical predictions.
Our results and findings are summarized in Sec.~\ref{sec:summary}. More details of the calculations are presented in the Appendices.
\section{Theory}\label{theory}
\subsection{Linearised Navier-Stokes equations under shear}
The Navier-Stokes equations of an isothermal MPC fluid are given by
\begin{align} \label{eq1}
\frac{\partial \rho}{\partial t} & = -\nabla\cdot(\rho \bf u) ,
 \\ \label{eq2}
\rho \left[ \frac{\partial}{\partial t} + \mathbf u \cdot \nabla \right]\mathbf u  & = -\nabla p + \eta \nabla^2
\mathbf u + \frac{\eta ^k}{3}\nabla \left (\nabla\cdot \mathbf u\right )~.
\end{align}
They account for mass and momentum conservation, where $\rho(\bf x,t) $ is the mass density, $\bf u(\bf x,t)$ the fluid velocity field, and $p(\bf x,t)$ the pressure field at the position $\bf x$  at time $t$. The shear viscosity is denoted as $\eta$. The Navier-Stokes equations are adopted to a non-angular-momentum-conserving MPC fluid, hence, the kinetic contribution $\eta^k$ of the shear viscosity appears in the last term in the rhs of Eq.~(\ref{eq2}), rather than the viscosity $\eta$ itself~\cite{huang2012}.
In addition, we omit the fluctuating part of the stress tensor~\cite{lutsko1985,otsuki2009-pre} in Eq.~(\ref{eq2}).
The equations are then linearised by setting $\rho=\rho_0+\delta \rho$, $p=p_0+\delta p$, and $\mathbf u=\mathbf u_0+\delta \mathbf u$, where $u_{0\alpha}=\gamma_{\alpha\beta} x_\beta$, with the shear-rate tensor $\gamma_{\alpha \beta}$ and $\alpha,  \beta \in \{x,y,z\}$.
We choose the $x$- and $y$-axis of the Cartesian coordinate system as the flow and the gradient direction, respectively,
such that $\gamma_{\alpha\beta}=\dot \gamma \delta_{\alpha x}\delta_{\beta y}$, where $\dot \gamma$ is the shear rate.
We use the summation convention for Greek indices unless otherwise stated.
Equations ~(\ref{eq1}) and (\ref{eq2}) can then be written as
\begin{align}
\left[\frac{\partial}{\partial t}+\gamma_{\alpha\beta}x_\beta\frac{\partial }{\partial x_\alpha}\right]\delta \rho  = & -\rho_0 \nabla\cdot \delta \mathbf u
\label{eq3} \\
\rho_0 \left[ \frac{\partial}{\partial t} + \gamma_{\alpha'\beta}x_\beta\frac{\partial}{\partial x_{\alpha'}}\right]\delta u_\alpha
=&-\rho_0 \gamma_{\alpha\beta}\delta u_\beta-\frac{\partial }{\partial x_\alpha}\delta p
\nonumber \\  + \eta \nabla^2 \delta u_\alpha
& + \frac{\eta ^k}{3}\frac{\partial}{\partial x_\alpha}\left (\nabla\cdot \delta \mathbf u \right)~.
\label{eq4}
\end{align}
Here, we have neglected second order terms in the fluctuations.  We  eliminate $\delta p$  with the ideal gas equation of state,
$\delta p=c_T^{2}\delta \rho$, where  $c_T$ is the isothermal velocity of sound. By  rescaling the velocity and density according to $\delta \mathbf u\equiv \delta \mathbf u/c_T$
and $\delta \rho\equiv \delta \rho/\rho_0$, Eqs.~(\ref{eq3}) and (\ref{eq4}) can be written in momentum space as
\begin{align}
\left[ \frac{\partial}{\partial t}-\gamma_{\alpha\beta}k_\alpha \frac{\partial}{\partial k_\beta}\right]\delta \tilde{\rho}  = & i c_T \mathbf k \cdot \delta \tilde{\mathbf u}
\label{eq8} \\
\left[ \frac{\partial}{\partial t}-\gamma_{\alpha'\beta}k_{\alpha'}\frac{\partial}{\partial k_\beta}\right]\delta \tilde{u}_\alpha
= &  -\gamma_{\alpha\beta}\delta \tilde u_\beta  +i c_T k_\alpha \delta \tilde{\rho} \nonumber \\
& - \nu k^2 \delta \tilde{u}_\alpha -\frac{\nu^k}{3} k_\alpha k_\beta~\delta \tilde{u}_\beta ~,
\label{eq9}
\end{align}
with the kinematic viscosities $\nu=\eta/\rho_0$, $\nu^k=\eta^k/\rho_0$. The variables with a tilde are Fourier-transformed variables according to the definition
\begin{equation}
\tilde{\mathbf f}\left(\mathbf k\right)=\int d^3 \mathbf x e^{i {\mathbf k \cdot \mathbf x} }\mathbf f(\mathbf x) .
\label{eq7}
\end{equation}

We now write the above equations in terms of the longitudinal and transverse component of the velocity field.
Let $\delta \tilde{\mathbf u}=\delta \tilde u^{\left(1\right)} {\mathbf e}^{\left(1\right)}+\delta \tilde u^{\left(2\right)} {\mathbf e}^{\left(2\right)}
+\delta \tilde u^{\left(3\right)} {\mathbf e}^{\left(3\right)}$,
where $ {\mathbf e}^{\left(1\right)}$,  ${\mathbf e}^{\left(2\right)}$, and ${\mathbf e}^{\left(3\right)}$ are three orthogonal unit vectors.
Here, ${\mathbf e}^{\left(1\right)}$ is chosen along the propagation direction of $\hat{\mathbf k}$, so that $\delta \tilde u^{\left(1\right)}$ is the longitudinal,
and $\delta \tilde u^{\left(2\right)}$ and $\delta \tilde u^{\left(3\right)}$ are the transverse
component of the velocity field. By introducing the vector $\tilde{\bf z}=(\delta \tilde{\rho},\delta \tilde u^{\left(1\right)},\delta \tilde u^{\left(2\right)}, \delta \tilde u^{\left(3\right)})^T$,
the Navier-Stokes  equations can be written as
\begin{equation}
\left[ \frac{\partial}{\partial t}-\dot\gamma k_x\frac{\partial}{\partial k_y}\right] \tilde{\mathbf z}+\mathcal L \tilde{\mathbf z}=0 .
\label{eq10}
\end{equation}
The explicit form of the matrix $\mathcal L$ for the choice~\cite{lutsko1985}
\begin{equation}
\begin{aligned}
{\mathbf e}^{\left(1\right)}&=\mathbf k / |{\bf k}|\\
{\mathbf e}^{\left(2\right)}&=[\hat {\mathbf y}-{e}^{\left(1\right)}_y {\mathbf e}^{\left(1\right)}]/\hat{k}_{\perp}\\
{\mathbf e}^{\left(3\right)}&={\mathbf e}^{\left(1\right)}\times{\mathbf e}^{\left(2\right)}
\label{unit_vectors}
\end{aligned}
\end{equation}
of the unit vectors is given in Appendix~\ref{appendixA}. Here, $\hat{\mathbf y}$ is the unit vector along the $y$-axis
in the Cartesian coordinate system and  $\hat{k}_{\perp}=(k_x^2+k_z^2)^{1/2}/k$, where $k=|\mathbf k|$.
The solution to the above equation can be written as the linear combination
\begin{equation}
\tilde {\bf z} \left(\mathbf k,t\right)=\sum_{i=1}^4a^{\left(i\right)}(\mathbf k,t) \boldsymbol \xi^{\left(i\right)}(\mathbf k)
\label{linear_comb}
\end{equation}
of the eigenvectors $\boldsymbol \xi^{(i)}(\mathbf k)$ which satisfy the eigenvalue equation
\begin{equation}
[-\dot\gamma k_x\frac{\partial}{\partial k_y}+\mathcal L] \boldsymbol \xi^{\left(i\right)}(\mathbf k)=\lambda_i \boldsymbol \xi^{\left(i\right)}(\mathbf k).
\label{eigen_value_eqn}
\end{equation}
Let $\boldsymbol \eta^{(i)}(\mathbf k)$ be the corresponding left eigenvectors such that
\begin{align} \label{eq:ortho}
\sum_{l=1}^4 \eta^{\left(i\right)}_l \xi^{\left(j\right)}_l =\delta_{ij} .
\end{align}
The left and right eigenvectors and the eigenvalues can be calculated using perturbation theory~\cite{lutsko1985} and are given
in Appendix.~\ref{appendixB}. Inserting  $\tilde {\bf z}(\mathbf k,t)$ from Eq.~(\ref{linear_comb}) into Eq.~(\ref{eq10}) and using Eq.~(\ref{eigen_value_eqn})
together with the orthogonality condition in Eq.~(\ref{eq:ortho}), we obtain
\begin{equation}
\left (\frac{\partial }{\partial t}-\dot{\gamma}k_x\frac{\partial}{\partial k_y}+\lambda_i(\bf k)\right)a^{\left(i\right)}\left({\bf k},t\right)=0 .
\end{equation}
The solution of the above equation is given by
\begin{equation}
a^{\left(i\right)}(\mathbf k,t)=a^{\left(i\right)}(\mathbf k(-t),0) \exp\left( -\int_0^t d\tau \lambda_i(\mathbf k(-\tau)) \right) ,
\label{a_solution}
\end{equation}
where the time dependent $\mathbf k$ vector is defined as $\mathbf k(t)=(k_x,k_y-\dot{\gamma}t k_x,k_z)$.
Using Eqs.~(\ref{linear_comb}) and (\ref{a_solution}) and the relation $a^{\left(i\right)}(\mathbf k,0)=
\sum_{l=1}^4{\eta}^{\left(i\right)}_l (\mathbf k) \tilde  z_l(\mathbf k,0)$, we get
\begin{equation}
\tilde z_{i}(\mathbf k,t)= \sum_{j=1}^4 G_{i j}(\mathbf k,t) \tilde z_{j}(\mathbf k(-t),0) ,
\label{solution1}
\end{equation}
where the propagator $G_{ij}(\mathbf k,t)$ is defined as
\begin{equation}
G_{ij}(\mathbf k,t)=\sum_{l=1}^{4}\xi^{(l)}_i(\mathbf k)\eta^{(l)}_j(\mathbf k(-t))\exp \left( -\int_0^t d\tau \lambda_l(\mathbf k(-\tau))\right) .
\label{propogator}
\end{equation}
In order to compare with the simulations, it is convenient to rewrite Eq.~(\ref{solution1}) by setting $\mathbf k=\mathbf k(t)$. We then get
\begin{equation}
\tilde z_i(\mathbf k(t),t)= \sum_{j=1}^4G_{ij}(\mathbf k(t),t)~\tilde z_{j}(\mathbf k,0) ,
\label{solution}
\end{equation}
with
\begin{equation}
G_{ij}(\mathbf k(t),t)=\sum_{l=1}^{4}\xi^{\left(l\right)}_i(\mathbf k(t))\eta^{\left(l\right)}_j(\mathbf k)\exp\left( -\int_0^t d\tau \lambda_l(\mathbf k(\tau)) \right) ,
\label{propogator_1}
\end{equation}
using $\left[\int_0^t d\tau \lambda_i(\mathbf k(-\tau))\right]_{\mathbf k=\mathbf k(t)}=\int_0^t d\tau \lambda_i(\mathbf k(\tau))$.
The explicit form of $G_{ij}(\mathbf k(t),t)$ can be obtained from the eigenvectors $\{\boldsymbol \xi,\boldsymbol \eta \}$ and the eigenvalues
$\lambda$'s given in Appendix~\ref{appendixB}.
Note that the solution given by Eq.~(\ref{solution}) represents the evolution of the hydrodynamic variables in the time-dependent reference frame in the $\mathbf k$-space.

\subsection{Hydrodynamic correlation functions}
The correlations of the hydrodynamic variables are defined as $C_{ij}(\mathbf k,\mathbf k',t)=\langle \tilde z_{i}(\mathbf k(t),t)\tilde z_{j}(\mathbf k',0)\rangle$, and become with Eq.~(\ref{solution})
\begin{equation}
C_{ij}(\mathbf k,\mathbf k' ,t)= \sum_{l=1}^4G_{il}(\mathbf k(t),t)~\langle \tilde z_{l}(\mathbf k,0)\tilde z_{j}(\mathbf k',0) \rangle.
\end{equation}
The correlations $\langle \tilde z_{i}(\mathbf k,0)\tilde z_{j}(-\mathbf k,0\rangle)$ vanishes at equilibrium, i.e., $\dot\gamma=0$, for $i \ne j$.
However, they are nonzero for $\dot\gamma\ne0$.
We consider only small  shear rates $\dot\gamma\lesssim\nu k^2$, for which
the cross-correlations can be neglected.
Hence, the correlation functions can be written as
$C_{ij}(\mathbf k,\mathbf k' ,t)\simeq(2\pi)^3\delta_{ij}\delta(\mathbf k +\mathbf k' )C_{ii}(\mathbf k,t)$, where $C_{ii}(\mathbf k,t)=
\langle \tilde z_{i}(\mathbf k,0)\tilde z_{i}(-\mathbf k,0)\rangle~G_{ii}(\mathbf k (t),t)$. Using the explicit
expressions for the propagators $G_{ii}(\mathbf k(t),t)$, the correlation functions can be written as
\begin{align}
C_{11}(\mathbf k,t)&=\frac{\rho_0 k_BT}{c_{T}^2}\left(\frac{k(t)}{k}\right)^{1/2}e^{-\frac{1}{2}\tilde \nu \chi(\mathbf k,t)}\cos\left[c_T\phi\left(\mathbf k,t\right)\right]\label{c_11} ,\\
C_{22}(\mathbf k,t)&=\frac{c_T^2}{\rho_0^2}C_{11}(\mathbf k,t)\label{c_22} ,\\
C_{33}(\mathbf k,t)&=\frac{k_BT}{\rho_0}\left(\frac{k}{k(t)}\right)e^{-\nu \chi(\mathbf k,t)} ,\label{c_33}\\
C_{44}(\mathbf k,t)&=\frac{k_BT}{\rho_0}e^{-\nu \chi(\mathbf k,t)} , \label{c_44}
\end{align}
where $\phi(\mathbf k,t)$ and $\chi(\mathbf k,t)$ are given by
\begin{align}
\phi(\mathbf k,t)&=\frac{1}{2\dot{\gamma}k_x}\Big[\left[k_yk-k_y(t)k(t)\right]\nonumber \\
&-k_{\perp}^2\ln\left(\frac{k_y(t)+k(t)}{k_y+k}\right)\Big] , \\
\chi(\mathbf k,t)&=k^2t-\dot{\gamma}k_xk_yt^2+\frac{1}{3}\dot{\gamma}^2k_x^2t^3 ,\label{chi}
\end{align}
and $k(t)=|\mathbf k(t)|$.
Here, $\tilde \nu=\nu+\nu^k/3$, and the equilibrium relations $\langle\tilde z_{1}\left(\mathbf k,0\right)\tilde z_{1}\left(-\mathbf k,0\right)\rangle=\rho_0k_BTc_{T}^{-2}$ and
$\langle \tilde z_{i}(\mathbf k,0)\tilde z_{i}(-\mathbf k,0)\rangle=\rho_0^{-1}k_BT$ for $i=2,3,4$ have been employed.
These expressions can be derived using fluctuating hydrodynamics for a MPC fluid \cite{huan:13}, however we do not present the derivations here.

A few remarks on the correlation functions given by Eqs.~(\ref{c_11})-(\ref{c_44}) are in order.
In the limit $\dot\gamma\rightarrow 0$, we get $\phi(\mathbf k,t)\rightarrow kt$ and $\chi(\mathbf k,t)\rightarrow k^2t$,
and therefore the correlation functions are reduced to the  corresponding equilibrium relations~\cite{hansen1990,huang2012} to $\mathcal O(k^2)$.
In the absence of shear, the correlation functions for an isothermal MPC fluid  can be obtained for all orders in $k$;
the exact expressions for the velocity autocorrelations are provided in Ref.~\cite{huang2012}.
We also note that the expression for $C_{33}(\mathbf k,t)$ remains exact for all shear rates within the
order we are working at, even if the neglected equal-time correlations of the form
$\langle \tilde z_{i}(\mathbf k,0)\tilde z_{j}(-\mathbf k,0)\rangle $ for $i\neq j$ are taken into account.
By the same token, $C_{44}({\bf k},t)$ is exact for all shear rates
for $k_z=0$.
\subsection{Velocity correlations in real space}
From Eq.~(\ref{solution1}), the velocity correlation function follows as
\begin{equation}
\langle \delta \tilde{\mathbf u}(\mathbf k,t) \cdot \delta \tilde{\mathbf u}(\mathbf k',0)\rangle=(2\pi)^3\delta(\mathbf k(-t)+\mathbf k')C^u(\mathbf k,t) ,
\end{equation}
with the abbreviation
\begin{equation}
C^u(\mathbf k,t)=\sum_{i=2}^{4}C_{ii}(\mathbf k(-t),t)~\mathbf e^{\left(i\right)}(\mathbf k)\cdot \mathbf e^{\left(i\right)}(\mathbf k(-t))
\end{equation}
and by using $C_{ij}(\mathbf k,t)\simeq 0$ for $i\neq j $.
The velocity autocorrelation is real space is then given by
\begin{equation}
\langle \delta \mathbf u(\mathbf x, t) \cdot\delta \mathbf u(\mathbf 0, 0)\rangle=
\frac{1}{\left(2\pi\right)^3}\int d^3\mathbf k~C^u(\mathbf k,t)e^{-i\mathbf k\cdot\mathbf x} .
\end{equation}
The velocity autocorrelation function $C(t)=\langle \mathbf v(t) \cdot \mathbf v(0)\rangle$ of a tagged particle of velocity ${\bf v}(t)$ can be obtained by setting ${\bf v}(t) = {\bf u}({\bf r},t)$, where $\bf r$ is the position of the tagged particle, and averaging over all its positions $\bf r$. Hence, we obtain
\begin{equation}
C(t)=\frac{1}{\left(2\pi\right)^3}\int d^3\mathbf k~C^u(\mathbf k,t)\langle e^{-i\mathbf k\cdot\mathbf r}\rangle~,
\label{c_t}
\end{equation}
with the definition $\langle e^{i\mathbf k\cdot\mathbf r}\rangle=\int d\mathbf r P(\mathbf r,t) e^{-i\mathbf k\cdot\mathbf r}$, and $P(\mathbf r,t)$ the distribution function
of the position of the tagged particle. Using the Fourier representation of $P(\mathbf r,t)$, we get $\langle e^{-i\mathbf k\cdot\mathbf r}\rangle=P(\mathbf k,t)$.
In shear flow, $P(\mathbf k,t)$ follows from the advective diffusion equation~\cite{dufty1984}
\begin{equation}
\left[\frac{\partial }{\partial t}-\dot\gamma k_x\frac{\partial}{\partial k_y}\right]P(\mathbf k,t)=-Dk^2P(\mathbf k,t) ,
\end{equation}
where $D$ is the diffusion coefficient. The solution of the equation is
\begin{equation}
P(\mathbf k,t)=P(\mathbf k(-t),0)\exp\left( {-D\int_0^td\tau k^2(-\tau)} \right) .
\end{equation}
Then, Eq.~(\ref{c_t}) yields
\begin{align}
C(t)=\frac{1}{\left(2\pi\right)^3}\int d^3 & \mathbf k \ C^u(\mathbf k,t)P(\mathbf k(-t),0)\\ & \times \exp\left({-D\int_0^td\tau k^2(-\tau)} \right) .
\end{align}
By changing the integration variable from $\mathbf k$ to $\mathbf k(t)$, and using the fact that the Jacobian of the transformation is unity, we get
\begin{align}
C(t)=\frac{1}{\left(2\pi\right)^3}\int d^3 & \mathbf k \ C^u(\mathbf k(t),t) \exp\left({-D\int_0^td\tau k^2(\tau)} \right)
\end{align}
by using $P(\mathbf k,0)=1$~\cite{ernst1971}.

So far, we considered infinitely large systems. In computer simulations, however, finite-size systems are used with typically periodic boundary conditions. This leads to a discrete set ${\bf k}_{\bf n}$ of wavevectors, with $k_{\alpha,n} = 2\pi n_{\alpha}/L$, where $L$ is the length of the cubic simulation box of volume $V= L^3$, $n_{\alpha} \in \mathbb{Z}$, and ${\bf k}_{\bf n} \ne 0$. Hence, the correlation function becomes
\begin{equation}
C(t)=\frac{1}{V}\sum_{\mathbf k_{\bf n}=- \infty}^ {\infty} C^u(\mathbf k_{\mathbf{n}}(t),t) \exp\left( -D\int_0^td\tau~\mathbf{k}_{\mathbf{n}}^2(\tau) \right) .
\end{equation}
The velocity autocorrelation function in shear flow is anisotropic. Therefore,
we write the above equation in terms of the components corresponds to the three orthogonal directions as
\begin{equation}
C_\alpha(t)=\frac{1}{V}\sum_{{\mathbf k}_{\bf n}=-\infty}^{\infty}\sum_{l=1}^{3} C_{jj}(\mathbf k_{\bf n},t)~e^{(l)}_{\alpha}(\mathbf k_{\bf n})  e^{(l)}_{\alpha}(\mathbf k_{\mathbf n}(t)) ,
\label{C_alpha}
\end{equation}
where $j=l+1$. Note that index $\alpha$ is not summed over.
Since MPC is a particle-based mesoscale simulation method, the validity of the Navier-Stokes equation breaks down at the level of collision cells~\cite{huang2012}.
Therefore the k-values in the summation in Eq.~(\ref{C_alpha}) are limited by a cut-off corresponding to the smallest hydrodynamic length scale.
%In writing the above equation, we have neglected the terms of the form $\langle \tilde %z_{\alpha}(k,0)\tilde z_{\beta}(-k,0)\rangle$ for $\alpha\ne\beta$.
Alternative but similar approaches to evaluate the velocity autocorrelations of a tagged fluid particle can be found in Refs.~\cite{otsuki2009-jsm,otsuki2009-epj}.

\section{Simulations} \label{sec:simulations}

\subsection{Multiparticle collision dynamics}
\begin{figure}
\includegraphics[width=\columnwidth]{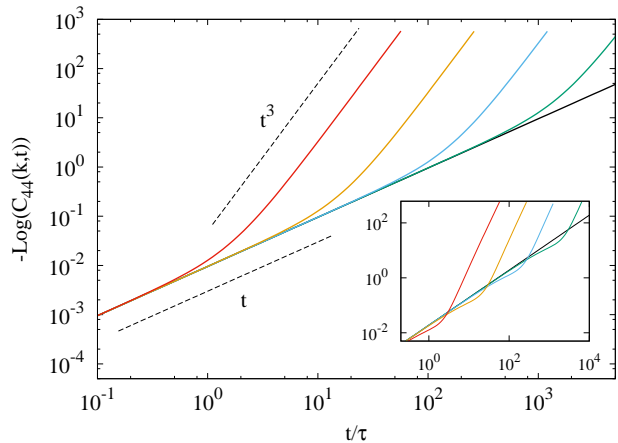}
\caption{(Color online) Numerically evaluated transverse velocity correlation function along
$\mathbf e^{\left(3\right)}(\mathbf k)$ (see Eqs.~(\ref{unit_vectors}) and (\ref{c_44})).
The lines (solid) correspond to shear rates~$\dot\gamma\tau=0.0,0.001,0.01,0.1,1.0$ (right to left).
In the main figure, the wavevector components are $k_x=2\pi/60$ and $k_y=k_z=0$
, and in the inset $k_x=k_y=2\pi/60$ and $k_z=0$.}
\label{trans_demonstration}
\end{figure}
In the MPC approach, the fluid is represented by point-particles \cite{kapr:08,gomp:09}. Their time evolution proceeds in two independent steps, namely the streaming and collision.
In the streaming step, the particles move ballistically, i.e., the particle positions are updated as
\begin{equation}
\mathbf x_i(t+h)=\mathbf x_i(t)+h\mathbf v_i(t)~,
\end{equation}
where $h$ is the collision-time step. Here, ${\bf x}_i$ denotes the position of particle $i$, ${\bf v}_i$ its velocity, and $i \in \{1,\ldots, N\}$, with the total number of particles $N$.  In the collision step, the particles are grouped into cubic cells of length $a$,
and a rotation of their relative velocities---with respect to the center-of-mass velocity of the particular cell---is performed. Hence, the new velocities are
\begin{equation}
\mathbf v_i(t+h)=\mathbf V_{cm}(t)+\mathcal R(\alpha)\left[\mathbf v_i(t)-\mathbf V_{cm}(t)\right]
\end{equation}
Here, $V_{cm}(t)$ is the center-of-mass velocity of the cell that contains the particle $i$ and $\mathcal R(\alpha)$ is the rotation matrix,
with the axis of rotation taken as a random unit vector. A random shift of the collision cell lattice is performed at every collision step to ensure
Galilean invariance \cite{ihle:01,gomp:09}.

We perform isothermal simulations, where temperature is maintained by the cell-level Maxwell-Boltzmann-Scaling (MBS) approach, which has been shown to yield a canonical ensemble \cite{huan:10.1,huan:15}.
The hydrodynamic fluctuations of the MPC fluid supplemented by the MBS method is known to be consistent with the linearized Navier-Stokes equation in equilibrium \cite{huang2012,huan:15}.
Shear flow is implemented by Lees-Edwards boundary conditions \cite{lees:72}.
The time step of our simulations is chosen as  $h/\tau =0.1$, with the unit of time $\tau=\sqrt{m a^2/k_BT}$, to ensure a large Schmidt number \cite{ripo:04}, and
the average number of particles in a collision cell is set to $10$. The numerical values of the transport coefficients for this choice of the simulation parameters
are $\nu=0.870 a^2/\tau$, $\tilde\nu=0.887 a^2/\tau$,  $D=0.051 a^2/\tau$, $c_T= 1.0 a/\tau$ \cite{gomp:09}.

\subsection{Hydrodynamic correlations}

\begin{figure}[t]
\includegraphics[width=\columnwidth]{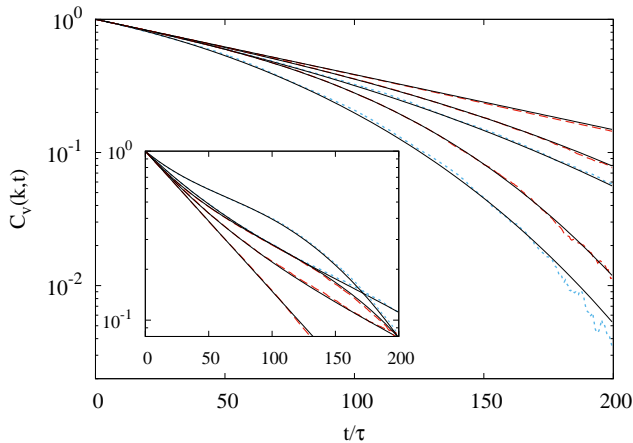}
\caption{(Color online) Theoretical and simulation results for the transverse velocity correlation functions along $\mathbf e^{\left(2\right)}(\mathbf k)$
(blue, dotted) and $\mathbf e^{\left(3\right)}(\mathbf k)$ (red, dashed) for shear rates
$\dot\gamma \tau=0.0,\ 0.005,\ 0.01$ (top to bottom at $t/\tau =100$). The two transverse components are identical for $\dot\gamma\tau=0.0$,
and therefore only one of them (red, dashed) is presented. The solid lines (black) represent the theoretical results.
In the main figure, the wavevector components are  $k_x=2\pi/L$ and $k_y=k_z=0$, and in the inset  $k_x=k_y=2\pi/L$ and $k_z=0$.
}
\label{cv_transverse}
\end{figure}
The density and velocity fields in $\bf k$-space are defined as
\begin{align}
\tilde{\rho}(\mathbf k,t) & = \sum_{i=1}^{N}e^{i\mathbf k(t).\mathbf x_i} , \\
\delta \tilde{\mathbf u}(\mathbf k,t) & = \sum_{i=1}^{N}\left[\mathbf v_{i}-\mathbf u_0(\mathbf x_i)\right]e^{i\mathbf k(t)\cdot\mathbf x_i},
\end{align}
where $\mathbf u_{0}(\mathbf x)=\dot\gamma y \hat{\mathbf x}$ is the mean velocity field.
Note that we use the time dependent ${\bf k}$-vector $\mathbf k(t)=(k_x,k_y-\dot\gamma t k_x,k_z)$,
so that the definitions of the hydrodynamic fields in momentum space are consistent with the Lees-Edwards boundary conditions.
The transverse and longitudinal components of the velocity field are then defined as
$\delta u^{\left(i\right)}(\mathbf k(t),t)= {\mathbf e}^{\left(i\right)}(\mathbf k(t))\cdot \delta \tilde{\mathbf u}(\mathbf k,t)$. \\

A few notes on the calculation of autocorrelation functions in shear flow implemented via Lees-Edwards boundary condition are in order.
In equilibrium simulations, the origin of time is arbitrary,  and therefore the moving-time-origin scheme~\cite{allen1989} for calculating
time correlation functions can be employed to improve statistics and to avoid storing position and velocity coordinates of the particles.
However, in our simulations, the $\bf k$-vector is taken as a function of time
and the time origin is taken as the time at which the image of a particle in the infinite periodic system is given
by $\mathbf x'_i=\mathbf x_i+\mathbf L$, where $\mathbf x_i$ is the position of  the particle in the primary simulation
box and $\mathbf L=L (n_x,n_y,n_z)^T$.
Therefore, averages have to be taken only over the allowed time origins.
In addition, in order to be consistent with the definition of the time-dependent $\bf k$-vector, the position coordinate in the
gradient direction has to be taken in the range $[-L_y/2,L_y/2]$.
However, the usual moving-time-origin scheme can be employed in the evaluation of real-space time-correlation functions.
\begin{figure}[t]
\includegraphics[width=\columnwidth]{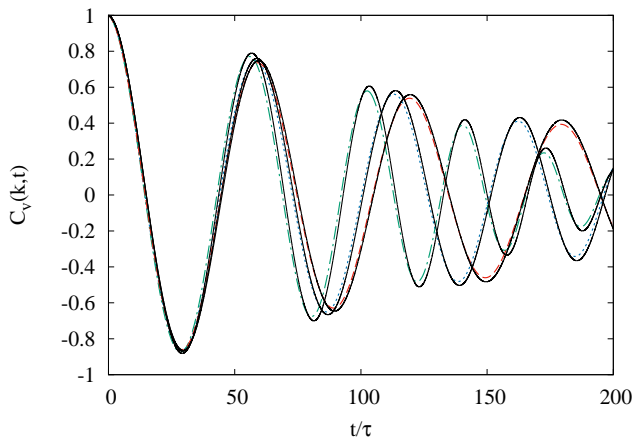}
\caption{(Color online) Longitudinal velocity correlations  for the shear rates
$\dot\gamma \tau=0.0$ (red, dashed) $0.005$ (blue, dotted), and $ 0.01$ (green, dotted-dashed). The solid lines (black) represent theoretical results.
The wavevector components are  $k_x=2\pi/L$ and $k_y=k_z=0$.}
\label{cv_longitudinal}
\end{figure}
\begin{figure}[b]
\includegraphics[width=\columnwidth]{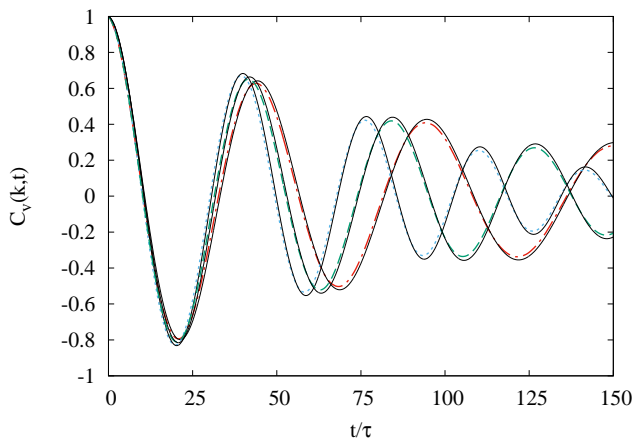}
\caption{(Color online) Longitudinal velocity correlations  for $\dot\gamma\tau=0.005$, $k_x=k_y=2\pi/L$ (dotted-dashed), $\dot\gamma\tau=0.005$, $k_x=-k_y=2\pi/L$ (dotted)
and $\dot\gamma\tau=0.0$, $k_x=k_y=2\pi/L$ (dashed). $k_z=0$ for all the curves. The solid (back) lines represent the theoretical results.}
\label{cv_doppler}
\end{figure}

\subsubsection{Correlation functions in momentum space}
\begin{figure*}
\includegraphics[width=0.9\linewidth]{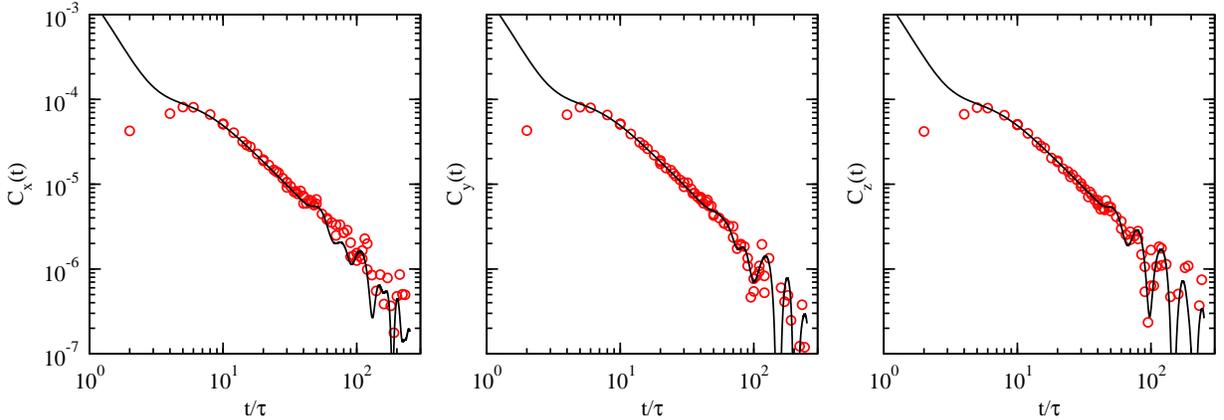}
\caption{(Color online) Velocity autocorrelation functions of a tagged particle along the various spatial directions. The shear rate is  $\dot\gamma \tau=0.01$.
Simulation results are represented by open circles and the theoretical prediction by solid lines.}
\label{cv_real}
\end{figure*}
Figure~\ref{trans_demonstration} shows the numerically evaluated transverse velocity correlation function given by Eq.~(\ref{c_44}).
As is evident from the theoretical expression, there are primarily two time-regimes for the decay of the correlations. For $t\ll 1/\dot\gamma$, the decay is dominated by the
term linear in $t$ in the exponential (see Eq.~(\ref{chi})) and therefore is identical to the decay of the correlation function in equilibrium. However, for $t\gg 1/\dot\gamma$, the decay
is dominated by the term proportional to $t^3$ and is characteristic of the shear flow.
This term originates from the advection term in the Navier-Stokes equation and results in several interesting features such as
faster decay with a power-law $ t^{-5/2}$ of the long-time tail in the velocity autocorrelation function of a tagged fluid particle~\cite{otsuki2009-jsm} and renormalization
of the viscosity~\cite{kumaran2006}.

In Fig.~\ref{cv_transverse} we compare the transverse velocity correlations obtained from the simulations and the theoretical expressions.
In contrast to equilibrium correlations, the autocorrelations of the two transverse components in shear flow are not identical.
The transverse velocity component  perpendicular ($\mathbf e^{(3)}$ direction) to the gradient direction decays slower
than the second component perpendicular ($\mathbf e^{(2)}$ direction) to the longitudinal direction for long times $(t\gg 1/\dot\gamma)$.
Even though the distinction between the two transverse components is apparent from our simulations,
it may be ignored in deriving the long-time tail exponents for the velocity autocorrelation function of a tagged fluid particle~\cite{otsuki2009-jsm,kumaran2006}.
The transverse correlations, as mentioned in the previous paragraph, decay similar to that in equilibrium for $t\ll 1/\dot\gamma$ and faster for $t\gg 1/\dot\gamma$.
For $t\approx 1/\dot\gamma$, the decay depends on the direction of the $\mathbf k$ vector, i.e., on the relative sign of $k_x$ and $k_y$. For $\text{sgn}(k_x)=\text{sgn}(k_y)$,
both the transverse correlations decay slower than the equilibrium correlations, and faster otherwise (see insets of Figs.~\ref{trans_demonstration} and ~\ref{cv_transverse}).
For $k_y=0$, the transverse correlation functions decay faster than the equilibrium correlations at all times.

The longitudinal velocity correlation function corresponds to the sound propagation in the fluid.
There are two effects of shear flow on the propagation of sound in an isothermal fluid -- the modification of the sound damping factor and the change in the
sound frequency/velocity (Doppler effect), both of which depend on the shear rate and the direction of propagation.
Figure~\ref{cv_longitudinal} shows the variation of longitudinal velocity correlations in the flow direction ($k_y=k_z=0$) for different shear rates.
The change in the frequency and the faster attenuation with increasing shear rate is well demonstrated.
Figure~\ref{cv_doppler} displays the anisotropy of the sound propagation. The frequency decreases when the sound propagation is in the direction along the flow, and
increases in the direction against the flow. The direction dependence of the attenuation of longitudinal velocity correlations is the same as that of the transverse velocity correlations.
The autocorrelation function of the density fluctuations shows an identical behavior as the longitudinal velocity correlations, and therefore we do present the results here.

\subsubsection{Long-time behavior of velocity correlations}

Figure~\ref{cv_real} shows velocity autocorrelation function of a tagged particle.
Note that we consider the thermal velocity of the particle, i.e., the velocity with respect to the mean flow velocity.
Evidently, the correlations in the three orthogonal directions are not identical.
We find excellent agreement between theory and simulation results for long times.
The deviations at short times are caused on the one hand by the fact that the theoretical hydrodynamic  correlations are only accurate to $\mathcal O(k^2)$. On the other hand, partition of the MPC fluid in collision cells leads to a break-down of hydrodynamics at short times and length scales below the collision-cell size \cite{huang2012}. However, the long-time behavior is determined by small $\bf k$ values, i.e., large length scales, which are correctly reproduced in the simulations.

\section{Summary and Conclusions} \label{sec:summary}

We have studied the nonequilibrium hydrodynamic time correlations of an isothermal MPC fluid under shear flow.
We find good agreement between simulation results and theoretical predictions based on the linearized Navier-Stokes equations for moderate shear rates.
We confirm that hydrodynamic correlations in shear flow are anisotropic, in agreement with previous studies~\cite{lutsko1985,otsuki2009-pre,otsuki2009-epj}.  Specifically and contrast to equilibrium correlations, the time correlations of the two transverse modes in Fourier space are no longer identical. In addition,
our simulations reveal the directional dependence of the frequency and attenuation of the longitudinal velocity correlation function. As a consequence, the velocity autocorrelation of a tracer fluid particle (MPC particle) is also anisotropic.
The agrement between analytical calculations and simulations confirms that MPC is a suitable approach to study hydrodynamic properties of simple fluids under nonequilibrium conditions.

Our studies are restricted to moderate shear rates, where equal-time correlations
of the hydrodynamic variables can be approximated by the corresponding equilibrium values.
For high shear rates, we observe significant deviations of the simulation results from the theoretical expressions.
The deviations increase with the shear rate.
In order to theoretically evaluate the equal-time and autocorrelation functions for high shear rates,
the fluctuating part of the stress tensor has to be included in the Navier-Stokes equations~\cite{lutsko1985},
which we omitted.
In addition, it is also necessary to take into account the density dependence of the viscosity in
linearising the Navier-Stokes equation. These issues will be addressed in future publications.

\appendix
\section{The hydrodynamic matrix}\label{appendixA}
The evolution of the hydrodynamic variables are given by
\begin{equation}
\left[ \frac{\partial}{\partial t}-\dot\gamma k_x\frac{\partial}{\partial k_y}\right] \tilde{\mathbf z}+\mathcal L\tilde {\mathbf z}=0~,
\end{equation}
where $\mathcal L=-ik\mathcal L_1+k^2 \mathcal L_2+\dot\gamma \mathcal L_3$, with
\begin{equation} \nonumber
\mathcal L_1=\left(
\begin{matrix}
0 && c_T && 0 && 0 \\
c_T && 0 && 0 && 0 \\
0 && 0 && 0 && 0 \\
0 && 0 && 0 && 0
\end{matrix}
\right)~,
\end{equation}
\begin{equation}
\mathcal L_2=\left(
\begin{matrix}
0 && 0 && 0 && 0 \\
0 && \tilde{\nu} && 0 && 0 \\
0 && 0 && \nu && 0 \\
0 && 0 && 0 && \nu
\end{matrix}
\right)~,
\end{equation}
\begin{equation}\nonumber
\mathcal L_3=\left(
\begin{matrix}
0 && 0 && 0 && 0 \\
0 && \Gamma_{11} && \Gamma_{12} && \Gamma_{13} \\
0 && \Gamma_{21} && \Gamma_{22} && \Gamma_{23} \\
0 && \Gamma_{31} && \Gamma_{32} && \Gamma_{33}
\end{matrix}
\right)~,
\end{equation}
where $\tilde{\nu}=\nu+\nu^k/3$, and the matrix $\boldsymbol \Gamma$ is defined as
\begin{equation}
\dot\gamma\Gamma_{ij}=e^{\left(i\right)}_m \gamma_{ml} e^{\left(j\right)}_l-e^{\left(i\right)}_n \gamma_{ml}k_m \frac{\partial }{\partial k_l} e^{\left(j\right)}_n
\end{equation}
For the particular choice of the unit vectors ${\mathbf e}^{\left(i\right)}$ as given in Eqs.~(\ref{unit_vectors}), the matrix $\boldsymbol \Gamma$ takes the form
\begin{equation}
\boldsymbol \Gamma=\left(
\begin{matrix}
k_xk_y/k^2 && 2k_xk_{\perp}/k^2 && 0 \\
-k_x/k_{\perp} && -k_xk_y/k^2&& 0\\
-k_yk_z/kk_{\perp} && -k_z/k && 0
\end{matrix}
\right)~,
\end{equation}
where $k_{\perp}^2=k_x^2+k_z^2$.
\newline
\section{Eigenvalues and eigenvectors of $\mathcal L$}\label{appendixB}
The eigenvalue equation Eq.~(\ref{eigen_value_eqn}) can be solved perturbatively by expanding $\boldsymbol \xi^{(m)}$ and $\lambda_m$ in powers of k
\begin{eqnarray}\nonumber
\boldsymbol \xi^{(m)}&=&\boldsymbol \xi^{(m)}_0+k\boldsymbol \xi^{(m)}_1+..\\
\lambda_m&=&k\lambda_{m,0}+k^2\lambda_{m,1}+..
\end{eqnarray}
The solution to the order $\mathcal O(k^2)$ is given by
\begin{equation}
\begin{aligned}
\lambda_1&=-ic_Tk+\frac{1}{2}\left(\tilde \nu k^{2}+\dot\gamma k_xk_y/k^2\right),\\
\lambda_2&=+ic_Tk+\frac{1}{2}\left(\tilde \nu k^{2}+\dot\gamma k_xk_y/k^2\right),\\
\lambda_3&=\nu k^2 -\dot\gamma k_xk_y/k^2, ~~ \lambda_4=\nu k^2,\\
\boldsymbol \xi^{(1)}&=\frac{1}{\sqrt{2}}(1,1,0,0)^T,~~\boldsymbol \xi^{(2)}=\frac{1}{\sqrt{2}}(1,-1,0,0)^T\\
\boldsymbol \xi^{(3)}&=(0,0,1,M)^T,~~\boldsymbol \xi^{(4)}=(0,0,0,1)^T~,
\end{aligned}
\label{lambda_def}
\end{equation}
where
\begin{equation}
M(\mathbf k)=-\frac{kk_z}{k_xk_\perp}\arctan\left(\frac{k_y}{k_\perp}\right)~.
\end{equation}
The left eigenvectors $\boldsymbol \eta^{(i)}$ which satisfy the condition $\sum_{l=1}^{4}\eta^{(i)}_l \xi^{(j)}_l=\delta_{ij}$ are given by
\begin{equation}
\boldsymbol \eta^{(m)}= \boldsymbol\xi^{(m)^T},~~\text{for}~~m=1,2,
\end{equation}
and
\begin{equation}
\boldsymbol \eta^{(3)}=(0,0,1,0),~~\boldsymbol \eta^{(4)}=(0,0,-M,1).
\end{equation}
%\bibliography{shear_flow_paper,bibliography_roland}
%\bibliography{Z:/Users/winkler/ownCloud/publications_library/bibliography}

\end{document}